\documentclass[conference,letterpaper]{IEEEtran}

\usepackage{xcolor}
\usepackage{graphicx}
\usepackage{amssymb}
\newtheorem{definition}{Definition}
\newtheorem{theorem}{Theorem}
\newtheorem{lemma}{Lemma}
\usepackage{comment}
\usepackage[ruled,vlined,linesnumbered]{algorithm2e}
\usepackage{algpseudocode}
\usepackage{balance}

\algrenewcommand\algorithmicrequire{\textbf{Input:}}
\algrenewcommand\algorithmicensure{\textbf{Output:}}

\newcounter{algnum}
\usepackage[ colorlinks = true,
linkcolor = blue,
urlcolor = blue,
citecolor = blue,
anchorcolor = green,
]{hyperref}

\usepackage[utf8]{inputenc} 
\usepackage[T1]{fontenc}
\usepackage{url}
\usepackage{ifthen}
\usepackage{cite}
\usepackage[cmex10]{amsmath} 

\begin{document}
\title{Reveal-or-Obscure: A Differentially Private Sampling Algorithm for Discrete Distributions} 

\author{%
  \IEEEauthorblockN{Naima Tasnim, Atefeh Gilani, Lalitha Sankar and Oliver Kosut}
  \IEEEauthorblockA{
                    Arizona State University,
                    Tempe, AZ, USA\\
                    Email: \{ntasnim2, agilani2, lsankar, okosut\}@asu.edu}
}

\maketitle



\begin{abstract}
    We introduce a differentially private (DP) algorithm called reveal-or-obscure (ROO) to generate a single representative sample from a dataset of $n$ observations drawn i.i.d. from an unknown discrete distribution $P$. Unlike methods that add explicit noise to the estimated empirical distribution, ROO achieves $\epsilon$-differential privacy by randomly choosing whether to ``reveal" or ``obscure" the empirical distribution. While ROO is structurally identical to Algorithm 1 proposed by Cheu and Nayak~\cite{cheu2024differentially}, we prove a strictly better bound on the sampling complexity than that extablished in Theorem 12 of~\cite{cheu2024differentially}. To further improve the privacy-utility trade-off, we propose a novel generalized sampling algorithm called Data-Specific ROO (DS-ROO), where the probability of obscuring the empirical distribution of the dataset is chosen adaptively. We prove that DS-ROO satisfies $\epsilon$-DP, and provide empirical evidence that DS-ROO can achieve better utility under the same privacy budget of vanilla ROO. 
\end{abstract}

\section{Introduction}
The widespread use of sensitive data across various domains, including healthcare, finance, law enforcement, and social sciences, has heightened the importance of privacy-preserving data analysis. 
Consequently, there is a growing need for mechanisms that allow data analysis while minimizing individual privacy risks. One promising approach is the use of synthetic data that capture the statistical properties of the original data.

Differential Privacy (DP)~\cite{dwork2006calibrating, dwork2014algorithmic} has emerged as a sound framework for formalizing privacy guarantees across a range of applications, including data analysis. In essence, DP ensures that the output of an algorithm does not differ by much whether or not an individual's data is included in the input. 

The task of synthetic data generation is closely related to the broader problem of learning probability distributions~\cite{kamath2019privately}. 
In the non-private setting, a learning algorithm approximates a distribution from which one can sample new data points that are representative of the original data. When privacy constraints are introduced, learning a distribution becomes significantly more challenging. In many practical cases, it may be sufficient to produce a small number of representative samples instead. The task of privately releasing one sample---known as DP sampling---is easier than full-fledged learning, since it requires less information from the underlying distribution. Motivated by this, we propose a novel DP sampling algorithm for discrete distributions on a finite alphabet. The key idea of our approach is to ``obscure'' the empirical distribution of the input dataset with a certain probability, or ``reveal'' it otherwise. Hence, we call our proposed algorithm reveal-or-obscure (ROO). 

\textbf{Main Contributions.} Our main contributions are:
\begin{itemize}
    \item We propose ROO---a sampling algorithm that achieves differential privacy without \emph{explicitly} perturbing the empirical distribution of the input dataset. We incorporate uncertainty in our algorithm by sampling from the uniform distribution with some fixed probability $q$.  
    \item We prove that our proposed algorithm reduces the sampling complexity while achieving better privacy-utility trade-off than the state-of-the art \cite{raskhodnikova2021differentially}, \cite{cheu2024differentially}.
    \item 
    We also propose DS-ROO (data-specific ROO) as a technique to generalize ROO by making $q$, i.e., the probability of sampling from the uniform distribution, a function of the empirical distribution of the dataset. We prove that it is possible to achieve the same privacy guarantee with a lower $q$ value relative to the vanilla ROO algorithm for sufficiently large datasets.
    \item  We demonstrate empirically that,  for the same privacy guarantee, DS-ROO achieves better utility than vanilla ROO as well as the state-of-the-art in~\cite{raskhodnikova2021differentially}. 
\end{itemize}

\textbf{Related Work.} 
The problem of differentially private sampling from unknown distributions is first investigated in~\cite{raskhodnikova2021differentially}. Raskhodnikova \textit{et al.}~\cite{raskhodnikova2021differentially} provide the first known bounds with $(\epsilon,\delta)$-DP guarantees on the complexity of sampling from arbitrary distributions over a discrete alphabet. DP algorithms for sampling from higher dimensional distributions such as multivariate Gaussians are presented in~\cite{ghazi2024differentially}. Husain \textit{et al.}~\cite{husain2020local} considers DP sampling in the local setting, where the central aggregator cannot be trusted and each user must produce a single data record privately. Private sampling has also been studied in the distributed setting \cite{acharya2020inference, acharya2020inference2}. A key focus of these efforts has been to reduce the sample complexity of DP-assured private sampling. 
While a recent and concurrent work~\cite{cheu2024differentially} independent from ours proposes an algorithm structurally identical to ROO, our analysis establishes a strictly better sampling complexity bound in the same setting. More generally, the problem of releasing a dataset in a differentially private manner has also been studied; for example, see \cite{bellovin2019privacy, hardt2012simple, zhu2017differentially, majeed2020anonymization, boedihardjo2022private}. Generating a single sample in a private manner is the first step towards releasing a larger synthetic dataset, and to this end, we focus on the former challenge in this paper.

\section{Problem Setup}
We begin by briefly reviewing some relevant definitions. We use uppercase letters, e.g., $X$, to denote random variables (RVs), and lowercase letters, e.g. $x$, for their instantiations. We assume that the dataset consists of $n$ RVs sampled from a finite alphabet of $k$ letters; without loss of generality, we take this alphabet to be $[k]= \{1, 2, \dots, k\}$. Let $\mathcal{P}$ be the class of probability distributions on $[k]$. Given a dataset $X^n = (X_1, X_2, \dots, X_n)$ of $n$ i.i.d. observations from some unknown $P \in \mathcal{P}$, a randomized algorithm (privacy mechanism) $\mathcal{A}: \mathcal{X}^n \mapsto \mathcal{X}$ outputs a single sample from $\mathcal{X}$. Let $Y = \mathcal{A}(X^n)$ be the random variable corresponding to the output of algorithm $\mathcal{A}$ given input $X^n$. The output $\mathcal{A}(X^n)$ is drawn from a distribution $Q$ such that
\begin{align}
    Q(y) = \sum_{x^n \in \mathcal{X}^n} \Pr \left\{ \mathcal{A}(x^n) = y | x^n \right\} \Pr \left\{ X^n = x^n \right\}.
\end{align}
The accuracy of $\mathcal{A}$ is measured by the \textit{closeness} between $Q$ and $P$. 
We use the \textit{total variation distance}, defined as
\begin{align}
    d_{TV} (Q, P) = \frac{1}{2} \|Q-P\|_1 = \frac{1}{2} \sum_x |Q(x) - P(x)|.
\end{align}
We use the following definition of sampling accuracy, introduced in \cite{axelrod2020sample}.
\begin{definition}[Accuracy of Sampling~\cite{axelrod2020sample}]\label{def:acc}
    An algorithm $\mathcal{A}$ is $\alpha$-accurate on a distribution $P$ if the total variation distance, $d_{TV}$ between $Q$ and $P$ is bounded by some constant $\alpha$, i.e.,
\begin{align}
    d_{TV} (Q, P) \leq \alpha. \label{eq:acc-dtv}
\end{align}
    An algorithm is $\alpha$-accurate on a class $\mathcal{P}$ of distributions if it is $\alpha$-accurate on every $P\in\mathcal{P}$.
\end{definition}
Two datasets $x^n$ and $\tilde{x}^n$ are considered \textit{neighbors}, denoted $x^n \sim \tilde{x}^n$, if they differ by at most one entry. DP is defined with respect to all such neighboring datasets as follows.   
\begin{definition}[Differential Privacy~\cite{dwork2006calibrating}]\label{def:dp}
    A randomized algorithm, or mechanism $\mathcal{A}: \mathcal{X}^n \to \mathcal{Y}$ is considered $\epsilon$-differentially private ($\epsilon$-DP) if, for every pair of neighboring datasets $x^n\sim\tilde{x}^n \in \mathcal{X}^n$, and for all $Y \subseteq \mathcal{Y}$, 
    \begin{align}
        \Pr\{ \mathcal{A}(x^n) \in Y \} \leq e^\epsilon \Pr\{ \mathcal{A}(\tilde{x}^n) \in Y \}.
    \end{align}
\end{definition}

In~\cite{raskhodnikova2021differentially}, the authors present an achievable $\epsilon$-DP sampler which does the following: \\
(i) computes, for each $j \in [k]$, the empirical probability distribution $\hat{p}_j$, \\ 
(ii) adds Laplace noise to each count, \\
(iii) uses an $L_1$ projection to restrict the Laplace-noised distribution to be a probability vector $\tilde{P} = (\tilde{p}_1, \dots, \tilde{p}_k)$, and\\
(iv) outputs an element of $[k]$ sampled from the distribution $\tilde{P}$. \\
For this algorithm, they show that for a sampling complexity 
\begin{align}\label{eq:n-prime}
    n'=\frac{2k}{\alpha \epsilon},
\end{align}
their algorithm is $\alpha$-accurate. In the following section, we prove that it is possible to achieve an $\epsilon$-DP and $\alpha$-accurate sampler with fewer samples than $n'$. Notably, for higher values of $\epsilon$, i.e., lower privacy, we gain an exponential reduction in the required number of samples. 
Moreover, \cite{raskhodnikova2021differentially} also establishes a lower bound on the sampling complexity as $\Omega\left(\frac{k}{\alpha \epsilon}\right)$ for a restricted range of $\epsilon \in (0,1]$. 

\section{Reveal-or-Obscure (ROO)}\label{sec:fixed-q}
Algorithm~\hyperref[alg:fixed-q-sampler]{1} presents our proposed private sampler ROO. We implement the idea of obscuring the empirical distribution $\hat{P}_{x^n}$ by sampling from the uniform distribution on $[k]$. However, we wish to do so with a small probability $q$, so that we do not deviate too much from the true distribution $P$. With probability $1-q$, we simply choose a sample from the given dataset, i.e., we reveal $\hat{P}_{x^n}$.

\vspace{5pt}
\phantomsection
\noindent\textbf{Algorithm \thealgnum:} Reveal-or-Obscure (ROO) \label{alg:fixed-q-sampler} 
\begin{algorithmic}[1]
\Require Dataset $x^n = (x_1, \dots, x_n)$, alphabet size $k$, privacy budget $\epsilon$, parameter $q$
\Ensure Sample $y$
\State With probability $q$, choose $y \sim \text{Unif}[1:k]$;
\State Otherwise, pick $i \sim \text{Unif}[1:n]$ and choose $y=x_i$;
\State \textbf{return} $y$;    
\end{algorithmic} 
\vspace{5pt}

The privacy and utility guarantees provided by Algorithm~\hyperref[alg:fixed-q-sampler]{1} is given by the following theorem. 
\begin{theorem}\label{th:fixed-q}
   Given $q$, Algorithm~\hyperref[alg:fixed-q-sampler]{1} is $\epsilon$-DP and $\alpha$-accurate for
   \begin{align}
       \epsilon &= \log \left( 1 + \frac{k(1-q)}{nq}\right), \text{ and} \label{eq:th1-eps} \\ 
       \alpha &= q \left( 1 - \frac{1}{k}\right), \label{eq:th1-acc}
   \end{align}
    from which we can solve for $q$ to obtain the sampling complexity as
   \begin{align}\label{eq:th1-n}
       n = \frac{k(1-\alpha)-1}{\alpha(e^\epsilon -1)}.
   \end{align}
\end{theorem}


\begin{lemma}\label{lm:sampl-comp}
For any $k \geq 2, \epsilon > 0$, and $\alpha \in \left(0, 1-\frac{1}{k}\right)$, the sampling complexity of Algorithm~\hyperref[alg:fixed-q-sampler]{1} is lower than that of ~\cite{raskhodnikova2021differentially} and~\cite{cheu2024differentially}.
\end{lemma}
The proofs of Theorem \ref{th:fixed-q} and Lemma \ref{lm:sampl-comp} are in Appendix~\ref{app:proof-th1} and~\ref{proof-sampl-comp}, respectively. 

In fact, the sampling complexity of Algorithm~\hyperref[alg:fixed-q-sampler]{1} is exponentially better in terms of $\epsilon$ compared to that of \cite{raskhodnikova2021differentially} in \eqref{eq:n-prime}. This would appear to violate their lower bound on the sampling complexity; the reason it does not is that their lower bound only applies for $\epsilon<1$.


\section{Data-Specific Reveal-or-Obscure (DS-ROO)}\label{sec:var-q}
For the ROO algorithm in Section~\ref{sec:fixed-q}, we fix $q$ in order to achieve $\epsilon$-DP for any possible dataset. From~\eqref{eq:ratio-final} in the privacy analysis of Appendix~\ref{app:proof-th1}, we observe that the supremum of ratio of probabilities is achieved by setting $p=0$. This corresponds to the case where one of two neighboring datasets under consideration is entirely missing an element of the alphabet, but this element is present in the other dataset. Thus, ROO is inefficient on datasets where each element of the alphabet appears reasonably often. In this section, we show that by making $q$ a function of the dataset---specifically, a function of the smallest empirical probability, the accuracy can be improved for the same privacy. Let $m \in \left\{0, 1, \dots, \left\lfloor \frac{n}{k} \right\rfloor\right\}$ denote the smallest number of times an element of $[k]$ appears in dataset $x^n$. Mathematically, $m$ can be expressed as $m = n \cdot \min_x \hat{P}_{x^n} (x)$. The modified private sampler---which we call the data-specific reveal-or-obscure (DS-ROO)---is presented in Algorithm~\hyperref[alg:var-q-sampler]{2}. In DS-ROO, the probability $q_m$ is determined from the value of $m$ associated with the given dataset.  
Notably, when $m=0$, the corresponding $q_0$ is equivalent to that of Algorithm~\hyperref[alg:fixed-q-sampler]{1}. That is, the case of a dataset that is missing an element of the alphabet (so $m=0$) is the worst-case scenario, and so in this case DS-ROO behaves identically to ROO. However, we will show empirically that in other cases, DS-ROO can do much better than vanilla ROO.

\vspace{5pt}
\phantomsection
\noindent\textbf{Algorithm 2:} Data-specific ROO (DS-ROO) \label{alg:var-q-sampler}
\begin{algorithmic}[1]
\Require Dataset $x^n = (x_1, \dots, x_n)$, alphabet size $k$, privacy budget $\epsilon$
\Ensure Sample $y$
\State Compute
\begin{align}
    q_0 &= \frac{1}{1 + \frac{n}{k} \left( e^\epsilon - 1 \right)}; \label{eq:q0}
\end{align}
\State \textbf{for} $m = 1,2, \dots, \left\lfloor \frac{n}{k} \right\rfloor$ \textbf{do}
\State \quad Compute
\begin{align}
    u_m &= - \frac{m}{n} + \frac{1}{k} - \frac{1}{n}, \label{eq:u-m}\\
    v_m &= e^{\epsilon} \left( \frac{1}{k} - \frac{m}{n} \right), \label{eq:v-m}\\
    w_m &= -\frac{1}{n} - \frac{m}{n} + \frac{m}{n} e^{\epsilon}, \label{eq:w-m}\\
    q_m &= \max\left\{0, \frac{u_m}{v_m} q_{m-1} - \frac{w_m}{v_m} \right\};\label{eq:qm}
\end{align}
\State \textbf{end for}
\State Compute $m = n \cdot \displaystyle\min_x \hat{P}_{x^n} (x)$;
\State With probability $q_m$, choose $y \sim \text{Unif}[1:k]$;
\State Otherwise, pick $i \sim \text{Unif}[1:n]$ and choose $y=x_i$;
\State \textbf{return} $y$;
\end{algorithmic}
\vspace{5pt}

\begin{figure}[htbp]
   \centering
   \includegraphics[width=0.41\textwidth]{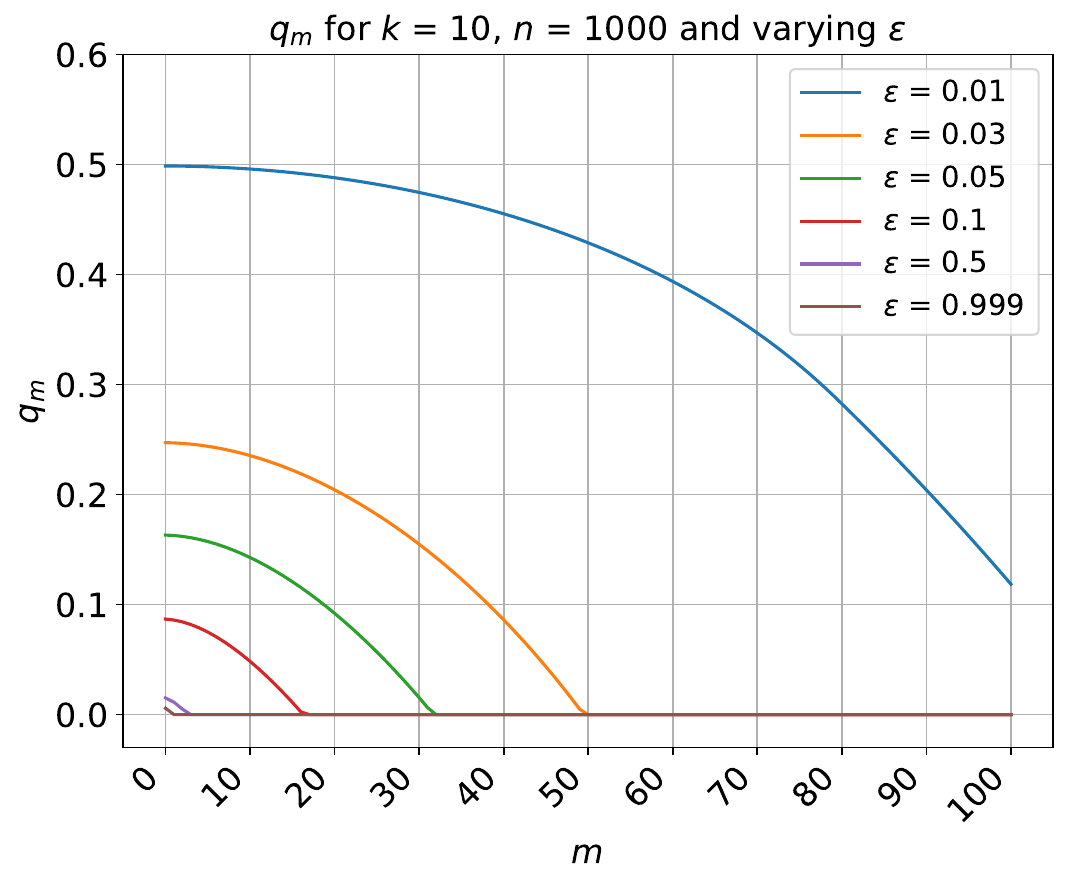}
  \caption{Plot of $q_m$ as a function of $m$ for fixed $k$ and $n$, showing changes under different privacy budgets $\epsilon$.}
   \label{fig:qM_var_eps_sec4}
 \end{figure}

Fig.~\ref{fig:qM_var_eps_sec4} shows the function $q_m$ for $k=9, n = 1000$, and several different values of the privacy parameter $\epsilon$. We observe that $q_m$ is non-increasing in $m$. As $m$ increases, the empirical distribution of the dataset gets closer to the uniform distribution, and $q_m$ approaches zero. Thus, DS-ROO is less likely to obscure the empirical distribution for larger $m$. In high privacy regimes, $q_m$ tends to decrease much slower. On the other hand, for larger $\epsilon$, i.e., low privacy, $q_m$ goes to zero much faster. The privacy guarantee provided by DS-ROO is given by the following theorem.

\begin{theorem}\label{th:var-q}
Algorithm~\hyperref[alg:var-q-sampler]{2} is $\epsilon$-differentially private.
\end{theorem}
The complete proof of Theorem~\ref{th:var-q} is provided in Appendix~\ref{app:proof-th2}. We present a proof sketch below.
\begin{IEEEproof}[Proof Sketch]
    Recall that $m$ denotes the smallest number of times an element of the alphabet $[k]$ appears in $x^n$. For neighboring datasets $x^n \sim \tilde{x}^n$, let $\tilde{m}$ denoted the $m$ value for $\tilde{x}^n$. Given $m$, there are three possible values of $\tilde{m}$: $(1)$ $\tilde{m} = m$, $(2)$ $\tilde{m} = m+1$, and $(3)$ $\tilde{m} = m-1$. Considering each of these cases separately and applying Definition~\ref{def:dp}, we derive the conditions under which DS-ROO satisfies $\epsilon$-DP as
\begin{gather}
    q_m \geq \frac{w_m}{u_m - v_m}, \text{ for } m = 0, 1, \ldots, 
    \left\lfloor\frac{1}{e^{\epsilon} - 1}\right\rfloor, \label{eq:qm-eq}\\
    u_m q_{m+1} \leq v_m q_m + w_m, \text{ for } m = 0, 1, \ldots, \left\lfloor\frac{n}{k}\right\rfloor - 1, \label{eq:qm-m-plus}\\
    u_m q_{m-1} \leq v_m q_m  + w_m, \text{ for } m = 1, 2, \ldots, \left\lfloor\frac{n}{k}\right\rfloor, \label{eq:qm-m-minus}
\end{gather}
    where $u_m, v_m$, and $w_m$ are as defined in~\eqref{eq:u-m}--\eqref{eq:w-m}. Additionally, for $q_m$ to be a valid probability, we must also have $0 \leq q_m \leq 1$ for all $m$. Setting the initial value $q_0$ as in~\eqref{eq:q0}, and assuming~\eqref{eq:qm-m-minus} to be an equality, we arrive at the expression
\begin{align}
   q_m = \max\left\{0, \frac{u_m}{v_m} q_{m-1} - \frac{w_m}{v_m} \right\}.
\end{align}
    In Lemmas~\hyperref[proof-tm]{3} and~\hyperref[proof-mplus1]{4} of Appendix~\ref{app:add-proofs}, we show that the above form of $q_m$ satisfies the inequality conditions in~\eqref{eq:qm-eq} and~\eqref{eq:qm-m-plus}, respectively. Therefore, Algorithm~\hyperref[alg:var-q-sampler]{2} is $\epsilon$-differentially private.
\end{IEEEproof}

\subsection{Utility of DS-ROO}
\begin{figure}[htbp]
   \centering
   \includegraphics[width=0.43\textwidth]{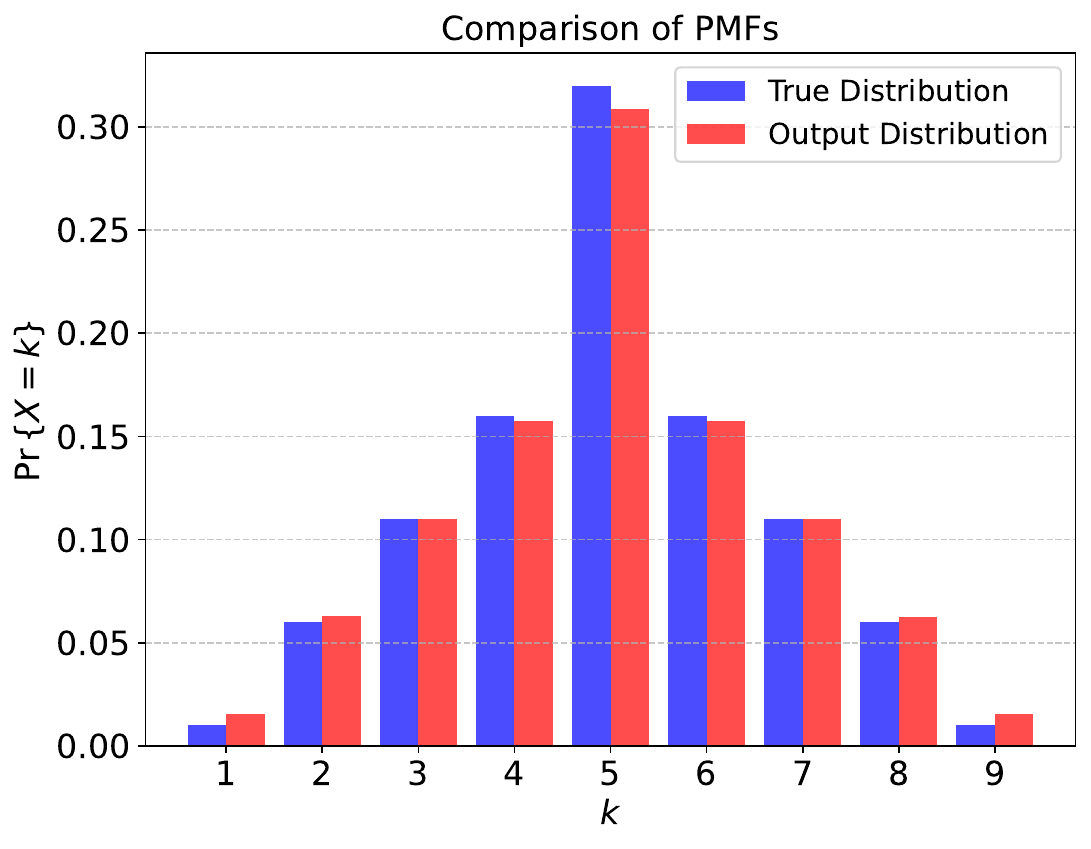}
  \caption{Comparison of true distribution and the estimated output distribution for DS-ROO.}
   \label{fig:comp-pmf}
 \end{figure}

We do not have theoretical bounds on the utility of DS-ROO at this time. However, we provide empirical evidence that DS-ROO achieves better utility than vanilla ROO and the state-of-the-art sampler in~\cite{raskhodnikova2021differentially} for the same privacy guarantee. In order to measure the utility of DS-ROO, we consider an input distribution, estimate the corresponding output distribution according to Algorithm~\hyperref[alg:var-q-sampler]{2}, and compute the total variation distance. Fig.~\ref{fig:comp-pmf} shows an example case for a distribution on an alphabet of size $k=9$, with dataset size $n=1000$, and privacy parameter $\epsilon=0.1$. We observe that mixing with the uniform distribution shifts some of the weight from the most probable central element to those with lower probabilities, resulting in reduced skewness in the output distribution. 

 \begin{figure}[htbp]
   \centering
   \includegraphics[width=0.43\textwidth]{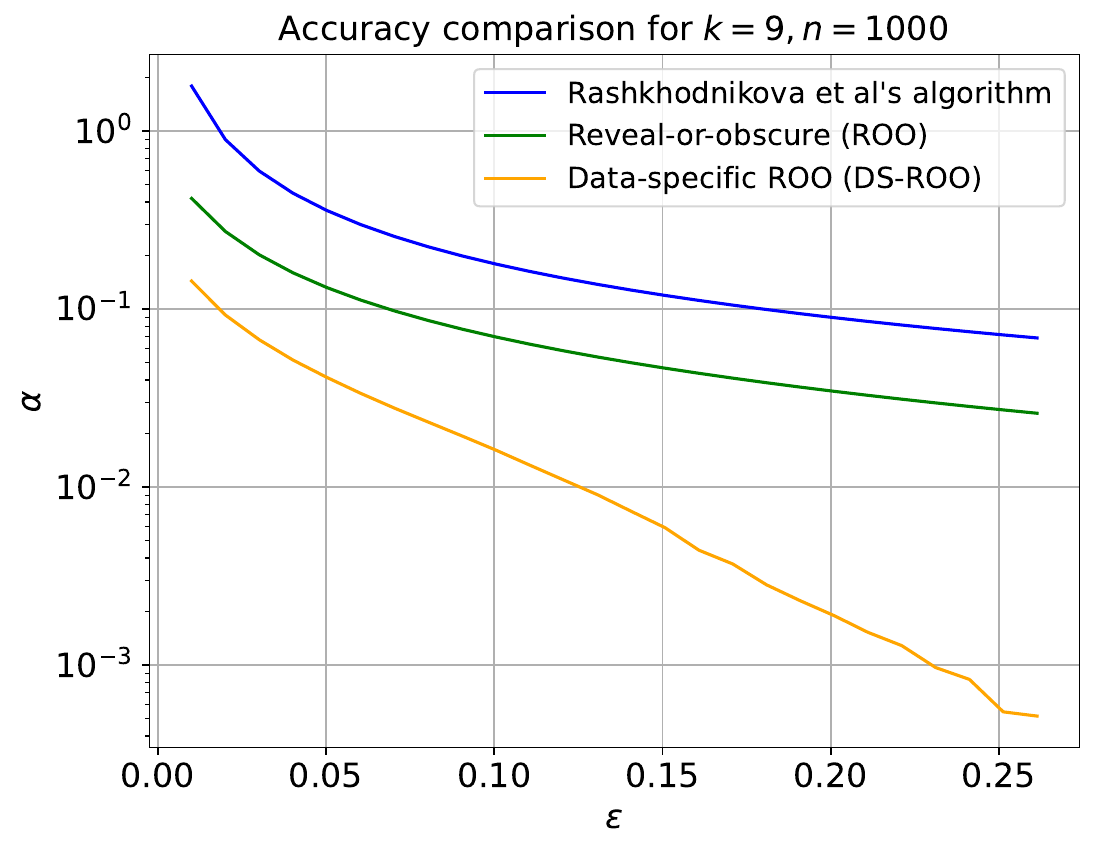}
  \caption{Comparison of accuracy ($\alpha$) versus privacy ($\epsilon$) curves of differentially private sampling algorithms.}
   \label{fig:comp-acc}
 \end{figure}

Recall from Definition~\ref{def:acc} that the total variation distance is upper bounded by $\alpha$. The smaller the value of $\alpha$, the more accurate the sampler is. From~\eqref{eq:n-prime}, the accuracy of the state-of-the-art sampler in~\cite{raskhodnikova2021differentially} is
\begin{align}
    \alpha = \frac{2k}{n' \epsilon}.
\end{align}
From Theorem~\ref{th:fixed-q}, we have the accuracy of ROO
\begin{align}
    \alpha = \frac{1}{1 + \frac{n}{k} (e^\epsilon - 1)} \left(1 - \frac{1}{k}\right).
\end{align}
For fixed values of $k$ and $n$, we obtain the accuracy of DS-ROO empirically, and compare it with that of~\cite{raskhodnikova2021differentially} and ROO. Fig.~\ref{fig:comp-acc} shows the $\alpha$ vs. $\epsilon$ curves for all three algorithms. We observe that DS-ROO achieves dramatically better accuracy than~\cite{raskhodnikova2021differentially} and vanilla ROO while providing the same privacy guarantee. Of course, the numerical results in Fig.~\ref{fig:comp-acc} are for the particular distribution shown in Fig.~\ref{fig:comp-pmf}. We anticipate that we will see similar improvements for distributions that are not too skewed---if the distribution is more skewed (such as if the probability of a letter is $0$), then there will be no improvement over vanilla ROO.

\section{Conclusion}
In this work, we propose a novel differentially private sampling algorithm for discrete distributions on a finite alphabet. Our algorithm achieves differential privacy by obscuring the empirical distribution of a dataset without perturbing it directly. In addition, we propose a method to generalize our approach to achieve better utility for the same privacy guarantee. For future work, we aim to explore the practicality of our proposed algorithms for more complex distribution classes.  

\appendix
\subsection{Proof of Theorem~\ref{th:fixed-q}}\label{app:proof-th1}
\textbf{Privacy Analysis.} Given $x^n$, Algorithm~\hyperref[alg:fixed-q-sampler]{1} chooses output $Y=y$ with probability 
\begin{align}\label{eq:emp-prob}
    P(Y=y|x^n) &= \frac{q}{k} + (1-q)\hat{P}_{x^n}(y),
\end{align}
where $\hat{P}_{x^n}(x)$ denotes the empirical probability of each $x \in [k]$, 
\begin{align}
    \hat{P}_{x^n}(x) = \frac{1}{n} \sum_{i=1}^n \mathbf{1}(x_i = x). 
\end{align}
For neighboring datasets $x^n \sim \tilde{x}^n$, their corresponding empirical probabilities of one observation differ by at most $\frac{1}{n}$, i.e., for all $x \in \mathcal{X}$,
\begin{align}\label{eq:neighbors}
    |\hat{P}_{x^n}(x) - \hat{P}_{\tilde{x}^n}(x)| \leq \frac{1}{n}.
\end{align}
By Definition~\ref{def:dp}, in order for ROO to satisfy $\epsilon$-DP, the following condition must hold for all possible $x^n \sim \tilde{x}^n$,
\begin{align}\label{eq:ratio}
    \frac{P(Y=y|x^n)}{P(Y=y|\tilde{x}^n)} \leq e^\epsilon.
\end{align}
We can write the left side of~\eqref{eq:ratio} as
\begin{align}
    \frac{P(Y=y|x^n)}{P(Y=y|\tilde{x}^n)} &= \frac{\frac{q}{k} + (1-q)\hat{P}_{x^n}(y)}{\frac{q}{k} + (1-q)\hat{P}_{\tilde{x}^n}(y)} \label{eq:ratio-sub} \\
    &\leq \sup_{ p } \frac{\frac{q}{k} + (1-q)\left( p+\frac{1}{n} \right)}{\frac{q}{k} + (1-q) p } \label{eq:ratio-sub-nb} \\
    &= \sup_{ p } 1 + \frac{(1-q)\frac{1}{n}}{\frac{q}{k} + (1-q) p } \\
    &\leq 1 + \frac{(1-q)\frac{1}{n}}{\frac{q}{k}} \label{eq:ratio-final}.
\end{align}
Here, in~\eqref{eq:ratio-sub}, we substitute~\eqref{eq:emp-prob}. In~\eqref{eq:ratio-sub-nb}, we use the property of the empirical probabilities stated in~\eqref{eq:neighbors}, denoting $\hat{P}_{\tilde{x}^n}(y)$ with $p$ for notational simplicity. In~\eqref{eq:ratio-final}, the supremum is obtained when $p=0$. 
Therefore, ROO satisfies $\epsilon$-DP guarantee if the right side of~\eqref{eq:ratio-final} is bounded by $\epsilon$, i.e.,
\begin{align}\label{eq:ratio-dp-upbound}
     1 + \frac{(1-q)\frac{1}{n}}{\frac{q}{k}} \leq e^\epsilon.
\end{align}
Rearranging \eqref{eq:ratio-dp-upbound}, we obtain the privacy guarantee of Theorem~\ref{th:fixed-q} stated in~\eqref{eq:th1-eps}.


\textbf{Utility Analysis.}
For our proposed sampler ROO in Algorithm~\hyperref[alg:fixed-q-sampler]{1}, the output distribution is
%
%
\begin{align}
    Q(y) &= \sum_{x^n \in \mathcal{X}^n} \left( \frac{q}{k} + (1-q) \hat{P}_{x^n} (y) \right) \Pr \left\{ X^n = x^n \right\}  \\
    &= \frac{q}{k} \sum_{x^n \in \mathcal{X}^n} \Pr \left\{ X^n = x^n \right\} \notag \\ &\quad + (1-q) \sum_{x^n \in \mathcal{X}^n} \hat{P}_{x^n} (y) \Pr \left\{ X^n = x^n \right\} \\
    &= \frac{q}{k} + (1-q) \mathbb{E}_{X^n} [\hat{P}_{x^n} (y)] \\
    &= \frac{q}{k} + (1-q)P(y), \label{eq:q-p}
\end{align}
where ~\eqref{eq:q-p} follows from the fact that $\hat{P}_{x^n}$ is the empirical distribution of a dataset sampled from $P$. 
The total variation distance between the discrete distributions $Q$ and $P$ is thus
\begin{align}
    d_{TV} (Q, P) &= \frac{1}{2} \sum_{y\in\mathcal{X}} |Q(y)-P(y)|  \\
    &= \frac{q}{2} \sum_{y\in\mathcal{X}} \left|\frac{1}{k} - P(y) \right| \label{eq:dtv-unif}  \\
    &= q \times  d_{TV}\left(\frac{1}{k}, P(y)\right) \\
    &\leq q \times \max_{P(y)} d_{TV}\left(\frac{1}{k}, P(y)\right), \label{eq:dtv-q-p}
\end{align}
where we obtain~\eqref{eq:dtv-unif} by rearranging and substituting~\eqref{eq:q-p}. Note that, for a convex objective function, the maximum is achieved at its corner points. Hence, the distribution $P$ that maximizes the $d_{TV}$ between the uniform distribution on $[k]$ and the input distribution $P$ must be one of the corner points of the $k$-dimensional probability simplex $\mathcal{P}$, e.g. $P = \{1, 0, \dots, 0\}$. The maximum $d_{TV}$ is then computed as
\begin{align}
    \max_{P(y)} d_{TV}\left(\frac{1}{k}, P(y)\right) &= \max_{P(y)} \frac{1}{2} \sum_{y\in\mathcal{X}} \left|\frac{1}{k} - P(y) \right| \\
    &= 1 - \frac{1}{k}.
\end{align}
Substituting this maximum objective value into~\eqref{eq:dtv-q-p}, we have
\begin{align}
    d_{TV} (Q, P) &\leq q \left( 1 - \frac{1}{k} \right).
\end{align}
Comparing this result with Definition~\ref{def:acc}, we obtain the utility of Theorem~\ref{th:fixed-q}, stated in~\eqref{eq:th1-acc}. We can then express $q$, i.e., the probability of sampling from the uniform distribution, in terms of $\alpha$ and $k$, 
\begin{align}
    q = \frac{\alpha}{1 - \frac{1}{k}} = \frac{k\alpha}{k - 1}.
\end{align}
Substituting the above into~\eqref{eq:ratio-dp-upbound} and rearranging, we obtain sample complexity 
\begin{align}
    n = \frac{k(1-\alpha) - 1}{\alpha(e^\epsilon - 1)}. 
    \label{eq:n-alpha}
\end{align}

\subsection{Proof of Lemma~\ref{lm:sampl-comp}}\label{proof-sampl-comp}
We first compare the sampling complexity of ROO with that of~\cite{raskhodnikova2021differentially}. From~\eqref{eq:th1-n} in Theorem~\ref{th:fixed-q}, we have the sampling complexity of ROO,
\begin{align}
    n = \frac{k(1-\alpha) - 1}{\alpha(e^\epsilon - 1)} &= \frac{k \left( 1 - \alpha - \frac{1}{k} \right) } {\alpha(e^\epsilon - 1)} \\
    &= \frac{2k}{\alpha \epsilon} \frac{\epsilon}{2(e^\epsilon - 1)} \left( 1 - \frac{1}{k} - \alpha \right) \\
    &< \frac{2k}{\alpha \epsilon} = n'.
\end{align}
As $\epsilon$ increases, the term $e^\epsilon - 1$ in the denominator of~\eqref{eq:th1-n} grows exponentially, resulting in significantly lower $n$ compared to $n'$.
Secondly, we compare the sampling complexity of ROO with that of~\cite{cheu2024differentially}. Theorem 12 of~\cite{cheu2024differentially} states that for any $\epsilon > 0$ and $\alpha \in (0,1)$, the SubRR algorithm is $\epsilon$-DP and performs $\alpha$-sampling for distributions over $[k]$ with sample complexity 
\begin{align}\label{eq:n-cheu}
    \frac{1}{\alpha \epsilon} (k - 1)(1 - \alpha) \leq \frac{k}{\alpha \epsilon}.
\end{align}
Subtracting the left side of~\eqref{eq:n-cheu} from~\eqref{eq:th1-n}, we have
\begin{align}
    &\frac{k(1-\alpha) - 1}{\alpha(e^\epsilon - 1)} - \frac{(k - 1)(1 - \alpha)}{\alpha \epsilon} \notag \\
    &\quad\quad= \frac{k(1 - \alpha) - 1 + \alpha - \alpha}{\alpha(e^\epsilon - 1)} - \frac{(k - 1)(1 - \alpha)}{\alpha \epsilon} \\
    &\quad\quad= \frac{(k-1)(1-\alpha) - \alpha}{\alpha(e^\epsilon - 1)} - \frac{(k - 1)(1 - \alpha)}{\alpha \epsilon} \\
    &\quad\quad= \frac{\epsilon(k-1)(1-\alpha) - \alpha\epsilon - (e^\epsilon - 1)(k-1)(1-\alpha)}{\alpha\epsilon(e^\epsilon - 1)} \\
    &\quad\quad= \frac{(\epsilon - e^\epsilon + 1)(k-1)(1-\alpha) - \alpha\epsilon}{\alpha\epsilon(e^\epsilon - 1)} \\
    &\quad\quad< 0, \label{eq:n-cheu-proof}
\end{align}
where~\eqref{eq:n-cheu-proof} follows from the fact that $1+\epsilon < e^\epsilon$ for $\epsilon > 0$. Therefore, despite the similar structure of ROO and SubRR, ROO requires fewer samples to achieve $\epsilon$-DP.

\subsection{Proof of Theorem~\ref{th:var-q}}\label{app:proof-th2}
Since $m$ denotes the smallest number of times an element in the alphabet $[k]$ appears in a dataset $x^n$, it is at most $\left\lfloor \frac{n}{k} \right\rfloor$ when the empirical distribution is uniform. For neighboring datasets $x^n \sim \tilde{x}^n$, there are three possible values of $\tilde{m}$: $(1)$ $\tilde{m} = m$, when the different entry in $\tilde{x}^n$ is a different element in $[k]$, hence it does not affect the minimum count, $(2)$ $\tilde{m} = m+1$, and $(3)$ $\tilde{m} = m-1$. In order to understand how the behavior of $q_m$ changes with $m$, we first analyze the likelihood ratio of the output distributions. 
Recall from Definition~\ref{def:dp} that, for Algorithm~\hyperref[alg:var-q-sampler]{2} to achieve $\epsilon$-DP, the following condition must hold for all possible $x^n \sim \tilde{x}^n$ pairs,
\begin{gather}
   \max_x \frac{\frac{q_{\tilde{m}}}{k} + (1-q_{\tilde{m}})\hat{P}_{\tilde{x}^n}(x)}{\frac{q_m}{k} + (1-q_m)\hat{P}_{x^n}(x)} \leq e^{\epsilon}.
\end{gather}
Adding and subtracting $\hat{P}_{x^n}(x)$ from $\hat{P}_{\tilde{x}^n}(x)$ in the numerator, the condition becomes
\begin{gather}
    \max_x \frac{\frac{q_{\tilde{m}}}{k} + (1-q_{\tilde{m}})(\hat{P}_{\tilde{x}^n}(x) - \hat{P}_{x^n}(x) + \hat{P}_{x^n}(x))}{\frac{q_m}{k} + (1-q_m)\hat{P}_{x^n}(x)} \leq e^{\epsilon}.
\end{gather}
The ratio is maximized when the numerator is maximized. Using~\eqref{eq:neighbors}, we can rewrite the condition as
\begin{gather}\label{eq:ratio-max}
    \frac{\frac{q_{\tilde{m}}}{k} + (1-q_{\tilde{m}})(\frac{1}{n} + \hat{P}_{x^n}(x))}{\frac{q_m}{k} + (1-q_m)\hat{P}_{x^n}(x)} \leq e^{\epsilon} \\
    \Rightarrow \hat{P}_{x^n}(x) \left( 1 - q_{\tilde{m}} - e^{\epsilon} (1-q_m) \right) \notag \\ \quad \leq e^{\epsilon} \frac{q_m}{k} - \frac{q_{\tilde{m}}}{k} - \left(1-q_{\tilde{m}}\right)\frac{1}{n}. \label{eq:ratio-rearr}
\end{gather}
Assuming $q_m$ and $q_{\tilde{m}}$ do not change significantly for $x^n \sim \tilde{x}^n$, we require
\begin{align}
    &1 - q_{\tilde{m}} < e^{\epsilon} (1-q_m) \Rightarrow
    1 - q_{\tilde{m}} - e^{\epsilon} (1-q_m) < 0.
\end{align}
Then. the inequality in~\eqref{eq:ratio-rearr} becomes
\begin{align}\label{eq:ratio-p}
    &\hat{P}_{x^n}(x) \geq \frac{ e^{\epsilon} \frac{q_m}{k} - \frac{q_{\tilde{m}}}{k} - (1-q_{\tilde{m}})\frac{1}{n} }{ 1 - q_{\tilde{m}} - e^{\epsilon} (1-q_m) }.
\end{align}
If~\eqref{eq:ratio-p} holds for all $x \in \mathcal{X}$, it must also hold for the minimum value of $\hat{P}_{x^n}(x)$, i.e.,
\begin{align}\label{eq:ratio-m}
    \frac{m}{n} \geq \frac{ e^{\epsilon} \frac{q_m}{k} - \frac{q_{\tilde{m}}}{k} - (1-q_{\tilde{m}})\frac{1}{n} }{ 1 - q_{\tilde{m}} - e^{\epsilon} (1-q_m) }.
\end{align}
Simplifying~\eqref{eq:ratio-m}, we arrive at the condition
\begin{align}\label{eq:qm-main}
    \left( - \frac{m}{n} + \frac{1}{k} - \frac{1}{n} \right) q_{\tilde{m}} &\leq e^{\epsilon} \left( \frac{1}{k} - \frac{m}{n} \right) q_m  - \frac{1}{n} - \frac{m}{n} + \frac{m}{n} e^{\epsilon}, \\
    \Rightarrow u_m q_{\tilde{m}} &\leq  v_m q_m + w_m,
\end{align}
where $u_m, v_m$ and $w_m$ are as defined in~\eqref{eq:u-m}--\eqref{eq:w-m}. 
Now, for $\tilde{m} = m$, we can replace $q_{\tilde{m}}$ with $q_m$ to obtain
\begin{align}\label{eq:qm-eq1}
    q_m \geq \frac{w_m}{u_m - v_m}.
\end{align}
Since the denominator is always negative, the right side of~\eqref{eq:qm-eq1} will be positive only when the numerator is also negative, i.e.,
\begin{align}
    w_m = -\frac{1}{n} - \frac{m}{n} + \frac{m}{n} e^{\epsilon} < 0 \Rightarrow m < \frac{1}{e^{\epsilon} - 1}.
\end{align}
For $\tilde{m} = m + 1$ and $\tilde{m} = m - 1$, we obtain two more inequalities,
\begin{align}\label{eq:qm-m-plus1}
    &u_m q_{m+1} \leq v_m q_m + w_m,
\end{align}
\begin{align}\label{eq:qm-m-minus1}
    &u_m q_{m-1} \leq v_m q_m  + w_m.
\end{align}
Therefore, for Algorithm~\hyperref[alg:var-q-sampler]{2} to achieve $\epsilon$-DP, the function $q_m$ must satisfy the inequalities in~\eqref{eq:qm-eq1},~\eqref{eq:qm-m-plus1}, and~\eqref{eq:qm-m-minus1}. Defining the initial value $q_0$ as in~\eqref{eq:q0}, and assuming~\eqref{eq:qm-m-minus1} to be an equality for $m \geq 1$, we arrive at the expression
\begin{align}\label{eq:qm-final}
    q_m = \frac{u_m}{v_m} q_{m-1} - \frac{w_m}{v_m}.
\end{align}
Note that, since $q_m$ is a probability, we must have $0 \leq q_m \leq 1$ for all $m$. Therefore,~\eqref{eq:qm-final} is valid only when the right side is positive. The final form of $q_m$ is thus
\begin{align}
   q_m = \max\left\{0, \frac{u_m}{v_m} q_{m-1} - \frac{w_m}{v_m} \right\}.
\end{align}
The proofs of~\eqref{eq:qm-final} satisfying the conditions in~\eqref{eq:qm-eq1} and~\eqref{eq:qm-m-plus1} are provided in Lemma~\ref{proof-tm} and~\ref{proof-mplus1}. The proof that the function $q_m$ is non-increasing is provided in Lemma~\ref{proof-qm-noninc}. When $m=0$, $q_m$ is equal to that of the fixed $q$ of Algorithm~\hyperref[alg:fixed-q-sampler]{1}. As $m$ increases, the input distribution gets closer to $\text{Unif}[1:k]$, and Algorithm~\hyperref[alg:var-q-sampler]{2} obscures the output distribution with smaller probability.  

\subsection{Additional Proofs}\label{app:add-proofs}
\begin{lemma}\label{proof-qm-noninc}
    For any $k \geq 2, \epsilon > 0$, and datasets of size $n$ where $n > k$, the function $q_m$ is non-increasing.
\end{lemma}
\begin{IEEEproof}
    Since the function $q_m: \left\{0, 1, \dots, \left\lfloor \frac{n}{k} \right\rfloor\right\} \mapsto [0,1]$ is discrete, we need to show that
        \begin{align}
            q_m \leq q_{m-1},
        \end{align}
        for all $m \in \left\{0, 1, \dots, \left\lfloor \frac{n}{k} \right\rfloor\right\}$. From Section~\ref{sec:var-q}, we have, for $m \geq 1$ and positive $q_m$,  
        \begin{align}
            v_m q_m &= u_m q_{m-1} - w_m \\
            \Rightarrow v_m q_m - v_m q_{m-1} &= u_m q_{m-1} - w_m - v_m q_{m-1} \\
            \Rightarrow q_m - q_{m-1} &= \frac{u_m - v_m}{v_m} q_{m-1} - \frac{w_m}{v_m} \\
            \Rightarrow q_m - q_{m-1} &= \left( \frac{u_m}{v_m} - 1 \right) q_{m-1} - \frac{w_m}{v_m}.\label{qjminusqjmin1}
        \end{align}
        It suffices to show that the right side of~\eqref{qjminusqjmin1} is negative. Substituting~\eqref{eq:u-m} and~\eqref{eq:v-m} into the first coefficient, we have
        \begin{align}
            \frac{u_m}{v_m} - 1 &= \frac{- \frac{m}{n} + \frac{1}{k} - \frac{1}{n}}{e^{\epsilon} \left( \frac{1}{k} - \frac{m}{n} \right)} - 1 \\
            &= \frac{\left(\frac{1}{k} - \frac{m}{n}\right) \frac{1}{n} - e^\epsilon \left(\frac{1}{k} - \frac{m}{n}\right)}{e^\epsilon \left(\frac{1}{k} - \frac{m}{n}\right) } \\
            &= \frac{-\frac{1}{n} - \left(\frac{1}{k} - \frac{m}{n}\right) (e^\epsilon-1)}{e^\epsilon \left(\frac{1}{k} - \frac{m}{n}\right) } \\
            &< 0.
        \end{align}
        Similarly, substituting~\eqref{eq:w-m} into the second coefficient, we have
        \begin{align}
            \frac{w_m}{v_m} &= \frac{-\frac{1}{n} - \frac{m}{n} + \frac{m}{n}e^\epsilon}{e^\epsilon\left(\frac{1}{k} - \frac{m}{n}\right) } \\ 
            &= \frac{-\frac{1}{n} + \frac{m}{n} (e^\epsilon-1)}{e^\epsilon\left(\frac{1}{k} - \frac{m}{n}\right)} \\
            &> 0, \label{wjvj}
        \end{align}
        for $m > \left\lfloor \frac{1}{e^\epsilon - 1} \right\rfloor$. Since $q_m$ has a lower bound $0$ in this range, the right side of~\eqref{qjminusqjmin1} is
        \begin{align}
            \left( \frac{u_m}{v_m} - 1 \right) q_{m-1} - \frac{w_m}{v_m} < 0.
        \end{align}
        It still remains to show that $q_m - q_{m-1} < 0$ for $m = 1, 2, \dots, \left\lfloor \frac{1}{e^\epsilon - 1} \right\rfloor$. Recall from~\eqref{eq:qm-eq1} in Appendix~\ref{app:proof-th2} that, in this range of $m$, $q_m$ has a non-zero, positive lower bound. Substituting this bound to the right side of \eqref{qjminusqjmin1}, we have
        \begin{align}
            &\left( \frac{u_m}{v_m} - 1 \right) \frac{w_{m-1}}{u_{m-1} - v_{m-1}} - \frac{w_m}{v_m} \notag \\
            &=  \frac{-\frac{1}{n} - \left(\frac{1}{k} - \frac{m}{n}\right) (e^\epsilon-1)}{e^\epsilon \left(\frac{1}{k} - \frac{m}{n}\right) } \frac{ \frac{m-1}{n} (e^{\epsilon} - 1) - \frac{1}{n} }{ \left(\frac{m-1}{n} - \frac{1}{k} \right) (e^{\epsilon} - 1) - \frac{1}{n} } - \frac{w_m}{v_m} \\
            &= \frac{-\frac{1}{n} - \left(\frac{1}{k} - \frac{m}{n}\right) (e^\epsilon-1)}{\left(\frac{m-1}{n} - \frac{1}{k} \right) (e^{\epsilon} - 1) - \frac{1}{n}} \frac{\frac{m-1}{n} (e^{\epsilon} - 1) - \frac{1}{n}}{e^\epsilon \left(\frac{1}{k} - \frac{m}{n}\right)}  - \frac{w_m}{v_m} \\
            &= C \frac{- \frac{1}{n} + \frac{m}{n} (e^{\epsilon} - 1) -  \frac{1}{n} (e^{\epsilon} - 1) }{e^\epsilon \left(\frac{1}{k} - \frac{m}{n}\right)}  - \frac{w_m}{v_m} \\
            &= C \left[ \frac{-\frac{1}{n} + \frac{m}{n} (e^\epsilon-1)}{e^\epsilon\left(\frac{1}{k} - \frac{m}{n}\right)} - \frac{\frac{1}{n}(e^\epsilon-1)}{e^\epsilon\left(\frac{1}{k} - \frac{m}{n}\right)} \right]  - \frac{w_m}{v_m} \\
            &= C \left[ \frac{w_m}{v_m} - \frac{e^\epsilon-1}{ e^\epsilon\left(\frac{n}{k} - m\right)} \right]  - \frac{w_m}{v_m}.
        \end{align}
        Here, $\frac{w_m}{v_m} < 0$ for the relevant range of $m$. Moreover, since $\frac{n}{k} > m$, $\frac{e^\epsilon-1}{ e^\epsilon\left(\frac{n}{k} - m\right)}$ is positive and less than $1$. Finally, the coefficient,
        \begin{align}
            C &= \frac{-\frac{1}{n} - \left(\frac{1}{k} - \frac{m}{n}\right) (e^\epsilon-1)}{\left(\frac{m-1}{n} - \frac{1}{k} \right) (e^{\epsilon} - 1) - \frac{1}{n}} \\
            &= \frac{-\frac{1}{n} - \left(\frac{1}{k} - \frac{m-1+1}{n}\right) (e^\epsilon-1)}{\left(\frac{m-1}{n} - \frac{1}{k} \right) (e^{\epsilon} - 1) - \frac{1}{n}} \\
            &= \frac{ \left(\frac{m-1}{n} - \frac{1}{k}\right) (e^\epsilon-1) -\frac{1}{n} + \frac{1}{n} (e^\epsilon-1)}{\left(\frac{m-1}{n} - \frac{1}{k} \right) (e^{\epsilon} - 1) - \frac{1}{n}} \\
            &= 1 + \frac{\frac{1}{n} (e^\epsilon-1)}{\left(\frac{m-1}{n} - \frac{1}{k} \right) (e^{\epsilon} - 1) - \frac{1}{n}} \\
            &= 1 - \frac{e^\epsilon-1}{1+n(e^\epsilon-1)\left(\frac{1}{k} -\frac{m-1}{n}\right) } \\
            &< 1.
        \end{align}
        Therefore, we have
        \begin{gather}
            C \left[ \frac{w_m}{v_m} - \frac{e^\epsilon-1}{ e^\epsilon\left(\frac{n}{k} - m\right)} \right] < \frac{w_m}{v_m} \\
            \Rightarrow C \left[ \frac{w_m}{v_m} - \frac{e^\epsilon-1}{ e^\epsilon\left(\frac{n}{k} - m\right)} \right] - \frac{w_m}{v_m} < 0
        \end{gather}
        Thus, for all $m$, 
        \begin{align}
            q_m \leq q_{m-1}, 
        \end{align}
        i.e., $q_m$ is non-increasing.
\end{IEEEproof}

\begin{lemma}\label{proof-tm}
    For $m = 1, 2, \dots, \left\lfloor \frac{1}{e^\epsilon - 1} \right\rfloor$, and $u_m, v_m$, 
    the function $q_m$ satisfies the inequality condition
    \begin{align}\label{eq:tm-proof}
        q_m > \frac{w_m}{u_m - v_m}.
    \end{align}
\end{lemma}
    \begin{IEEEproof}
        Let,
        \begin{align}
            t_m = \frac{w_m}{u_m - v_m} &= \frac{ \frac{w_m}{v_m} }{\frac{u_m}{v_m} - 1} \\
            \Rightarrow \left(\frac{u_m}{v_m} - 1\right) t_m &= \frac{w_m}{v_m}.  
        \end{align}
        We observe that
        \begin{align}
            t_0 = \frac{w_0}{u_0 - v_0} = \frac{-\frac{1}{n}}{\frac{1}{k}-\frac{1}{n}-\frac{e^\epsilon}{k}} = \frac{1}{1 + \frac{n}{k}(e^\epsilon-1)} = q_0.
        \end{align}
        Moreover, 
        \begin{align}
            t_m &= \frac{\frac{m}{n}(e^\epsilon - 1) - \frac{1}{n}}{\left(\frac{m}{n}-\frac{1}{k}\right)(e^\epsilon-1)-\frac{1}{n}} \\
            &= \frac{\frac{m}{n}(e^\epsilon - 1) - \frac{1}{n} - \frac{1}{k}(e^\epsilon-1) + \frac{1}{k}(e^\epsilon-1)}{\left(\frac{m}{n}-\frac{1}{k}\right)(e^\epsilon-1)-\frac{1}{n}} \\
            &= \frac{\left(\frac{m}{n}-\frac{1}{k}\right)(e^\epsilon-1)-\frac{1}{n} + \frac{1}{k}(e^\epsilon-1)}{\left(\frac{m}{n}-\frac{1}{k}\right)(e^\epsilon-1)-\frac{1}{n}} \\
            &= 1 + \frac{\frac{1}{k}(e^\epsilon-1)}{\left(\frac{m}{n}-\frac{1}{k}\right)(e^\epsilon-1)-\frac{1}{n}} \\
            &= 1 - \frac{\frac{1}{k}(e^\epsilon-1)}{\left(\frac{1}{k}-\frac{m}{n}\right)(e^\epsilon-1)+\frac{1}{n}}. \label{eq:t-m}
        \end{align}
        As $m$ increases,~\eqref{eq:t-m} becomes smaller due to a larger fraction being subtracted from $1$. Thus, $t_m$ is decreasing in $m$. Now, we prove the lemma for the base case and apply induction. 
        \textit{Base case:} For $m=1$, we have
        \begin{align}
            q_1 &= \frac{u_1}{v_1} q_0 - \frac{w_1}{v_1} \\
            &= \frac{u_1}{v_1} t_0 - \left(\frac{u_1}{v_1} - 1\right) t_1 \\
            &= \frac{u_1}{v_1} (t_0 - t_1) + t_1 \\
            &> t_1,
        \end{align}
        since $\frac{u_1}{v_1} > 0$, and $t_m$ is strictly decreasing in $m$. 
        
        \textit{Induction step:} Let us assume $q_m > t_m$ for any $m$. We have
        \begin{align}
            q_{m+1} &= \frac{u_{m+1}}{v_{m+1}} q_m - \frac{w_{m+1}}{v_{m+1}} \\
            &= \frac{u_{m+1}}{v_{m+1}} q_m - \left(\frac{u_{m+1}}{v_{m+1}} \right) t_{m+1} \\
            &> \frac{u_{m+1}}{v_{m+1}} t_m - \left(\frac{u_{m+1}}{v_{m+1}} \right) t_{m+1} \\
            &= \frac{u_{m+1}}{v_{m+1}} (t_m - t_{m+1}) + t_{m+1}  \\
            &> t_{m+1}.
        \end{align}
        Thus, for $m = 1, 2, \dots, \left\lfloor \frac{1}{e^\epsilon - 1} \right\rfloor$, $q_m$ satisfies~\eqref{eq:tm-proof}.
\end{IEEEproof}

\begin{lemma}\label{proof-mplus1}
    For $m \in \left\{0, 1, \dots, \left\lfloor \frac{n}{k} \right\rfloor - 1\right\}$, the function $q_m$ satisfies the inequality condition
    \begin{align}\label{eq:qm-plus-lemma}
        &u_m q_{m+1} < v_m q_m + w_m.
    \end{align}
\end{lemma}
    \begin{IEEEproof}
        Using~\eqref{eq:u-m}--\eqref{eq:w-m}, we have, for $m=0$,
        \begin{align}
            u_0 = \frac{1}{k} - \frac{1}{n}, v_0 = \frac{e^\epsilon}{k}, w_0 = - \frac{1}{n}.
        \end{align}
        Then, we can write 
        \begin{align}
            u_0 q_1 - v_0 q_0 &= u_0 q_1 - u_0 q_0 + u_0 q_0 - v_0 q_0 \notag \\
            &= u_0 (q_1 - q_0) + \left( \frac{1}{k} - \frac{1}{n} \right) q_0 - \frac{e^\epsilon}{k} q_0 \\
            &= u_0 (q_1 - q_0) + \left( \frac{1}{k} + w_0 \right) q_0 - \frac{e^\epsilon}{k} q_0 \\
            &= u_0 (q_1 - q_0) + \frac{1}{k} q_0 + w_0 q_0 - \frac{e^\epsilon}{k} q_0 \\
            &= w_0 q_0 - \frac{e^\epsilon-1}{k} q_0 - u_0 (q_0 - q_1) \\
            &< w_0. \label{eq:a}
        \end{align}
        Here,~\eqref{eq:a} follows from the fact that $0 \leq q_0 \leq 1$ and $q_m$ is non-increasing. 
        
        For $m=1, 2, \dots, \left\lfloor \frac{n}{k} \right\rfloor - 1$, we can write
        \begin{align}
            u_m q_{m+1} - v_m q_m &= u_m q_{m+1} - u_m q_{m-1} + w_m \label{eq:b} \\
            &= w_m + u_m (q_{m+1} - q_{m-1}) \\
            &= w_m + \left( - \frac{m}{n} + \frac{1}{k} - \frac{1}{n} \right) (q_{m+1} - q_{m-1}) \\
            &= w_m - \left(\frac{1}{k} - \frac{m+1}{n} \right) (q_{m-1} - q_{m+1}) \\
            &< w_m. \label{eq:c}
        \end{align}
        Here, in~\eqref{eq:b}, we substitute~\eqref{eq:qm-final}. \eqref{eq:c} follows from the fact that $q_m$ is non-increasing and $\frac{1}{k} - \frac{m+1}{n}$ is non-negative for $m=1, 2, \dots, \left\lfloor \frac{n}{k} \right\rfloor - 1$. Finally, for $q_{m+1} = 0$, we have
        \begin{align}
            - v_m q_m &= w_m - \left(\frac{1}{k} - \frac{m+1}{n} \right) q_{m-1} \\
            &< w_m \\
            \Rightarrow q_m &> - \frac{w_m}{v_m}. \label{d}
        \end{align} 
        It follows from~\eqref{wjvj} that the right side of~\eqref{d} is negative. Thus,~\eqref{d} is a valid inequality, and we can conclude that $q_m$ satisfies~\eqref{eq:qm-plus-lemma} for the relevant range of $m$.
    \end{IEEEproof}

\balance

\section*{Acknowledgment}
This work is supported in part by NSF grants CIF-1901243, CIF-2312666, and CIF-2007688.



\vfill
\bibliographystyle{IEEEtran}    
\bibliography{main}

\begin{thebibliography}{10}
\providecommand{\url}[1]{#1}
\csname url@samestyle\endcsname
\providecommand{\newblock}{\relax}
\providecommand{\bibinfo}[2]{#2}
\providecommand{\BIBentrySTDinterwordspacing}{\spaceskip=0pt\relax}
\providecommand{\BIBentryALTinterwordstretchfactor}{4}
\providecommand{\BIBentryALTinterwordspacing}{\spaceskip=\fontdimen2\font plus
\BIBentryALTinterwordstretchfactor\fontdimen3\font minus \fontdimen4\font\relax}
\providecommand{\BIBforeignlanguage}[2]{{%
\expandafter\ifx\csname l@#1\endcsname\relax
\typeout{** WARNING: IEEEtran.bst: No hyphenation pattern has been}%
\typeout{** loaded for the language `#1'. Using the pattern for}%
\typeout{** the default language instead.}%
\else
\language=\csname l@#1\endcsname
\fi
#2}}
\providecommand{\BIBdecl}{\relax}
\BIBdecl

\bibitem{cheu2024differentially}
A.~Cheu and D.~Nayak, ``Differentially private multi-sampling from distributions,'' \emph{arXiv preprint arXiv:2412.10512}, 2024.

\bibitem{dwork2006calibrating}
C.~Dwork, F.~McSherry, K.~Nissim, and A.~Smith, ``Calibrating noise to sensitivity in private data analysis,'' in \emph{Theory of Cryptography: Third Theory of Cryptography Conference, TCC 2006, New York, NY, USA, March 4-7, 2006. Proceedings 3}.\hskip 1em plus 0.5em minus 0.4em\relax Springer, 2006, pp. 265--284.

\bibitem{dwork2014algorithmic}
C.~Dwork, A.~Roth \emph{et~al.}, ``The algorithmic foundations of differential privacy,'' \emph{Foundations and Trends{\textregistered} in Theoretical Computer Science}, vol.~9, no. 3--4, pp. 211--407, 2014.

\bibitem{kamath2019privately}
G.~Kamath, J.~Li, V.~Singhal, and J.~Ullman, ``Privately learning high-dimensional distributions,'' in \emph{Conference on Learning Theory}.\hskip 1em plus 0.5em minus 0.4em\relax PMLR, 2019, pp. 1853--1902.

\bibitem{raskhodnikova2021differentially}
S.~Raskhodnikova, S.~Sivakumar, A.~Smith, and M.~Swanberg, ``Differentially private sampling from distributions,'' \emph{Advances in Neural Information Processing Systems}, vol.~34, pp. 28\,983--28\,994, 2021.

\bibitem{ghazi2024differentially}
B.~Ghazi, X.~Hu, R.~Kumar, and P.~Manurangsi, ``On differentially private sampling from gaussian and product distributions,'' \emph{Advances in Neural Information Processing Systems}, vol.~36, 2024.

\bibitem{husain2020local}
H.~Husain, B.~Balle, Z.~Cranko, and R.~Nock, ``Local differential privacy for sampling,'' in \emph{International Conference on Artificial Intelligence and Statistics}.\hskip 1em plus 0.5em minus 0.4em\relax PMLR, 2020, pp. 3404--3413.

\bibitem{acharya2020inference}
J.~Acharya, C.~L. Canonne, and H.~Tyagi, ``Inference under information constraints i: Lower bounds from chi-square contraction,'' \emph{IEEE Transactions on Information Theory}, vol.~66, no.~12, pp. 7835--7855, 2020.

\bibitem{acharya2020inference2}
------, ``Inference under information constraints ii: Communication constraints and shared randomness,'' \emph{IEEE Transactions on Information Theory}, vol.~66, no.~12, pp. 7856--7877, 2020.

\bibitem{bellovin2019privacy}
S.~M. Bellovin, P.~K. Dutta, and N.~Reitinger, ``Privacy and synthetic datasets,'' \emph{Stan. Tech. L. Rev.}, vol.~22, p.~1, 2019.

\bibitem{hardt2012simple}
M.~Hardt, K.~Ligett, and F.~McSherry, ``A simple and practical algorithm for differentially private data release,'' \emph{Advances in neural information processing systems}, vol.~25, 2012.

\bibitem{zhu2017differentially}
T.~Zhu, G.~Li, W.~Zhou, and S.~Y. Philip, ``Differentially private data publishing and analysis: A survey,'' \emph{IEEE Transactions on Knowledge and Data Engineering}, vol.~29, no.~8, pp. 1619--1638, 2017.

\bibitem{majeed2020anonymization}
A.~Majeed and S.~Lee, ``Anonymization techniques for privacy preserving data publishing: A comprehensive survey,'' \emph{IEEE access}, vol.~9, pp. 8512--8545, 2020.

\bibitem{boedihardjo2022private}
M.~Boedihardjo, T.~Strohmer, and R.~Vershynin, ``Private sampling: a noiseless approach for generating differentially private synthetic data,'' \emph{SIAM Journal on Mathematics of Data Science}, vol.~4, no.~3, pp. 1082--1115, 2022.

\bibitem{axelrod2020sample}
B.~Axelrod, S.~Garg, V.~Sharan, and G.~Valiant, ``Sample amplification: Increasing dataset size even when learning is impossible,'' in \emph{International Conference on Machine Learning}.\hskip 1em plus 0.5em minus 0.4em\relax PMLR, 2020, pp. 442--451.

\end{thebibliography}

\end{document}